\pdfoutput=1

\newif\iflockedminted
\lockedmintedtrue

\newif\ifarxiv
\arxivtrue

\ifarxiv
\documentclass[acmsmall,nonacm,screen]{acmart}
\else
\documentclass[acmsmall,review,anonymous]{acmart}
\fi

\ifarxiv
\setcopyright{cc}
\setcctype[4.0]{by}
\else
\renewcommand\footnotetextcopyrightpermission[1]{}
\fi

\usepackage{xcolor,soul}
\usepackage[noline,noend]{algorithm2e}
\usepackage{listings}
\usepackage[shortlabels]{enumitem}

\iflockedminted
\usepackage[frozencache=true,cachedir=minted-cache]{minted}
\else
\usepackage[finalizecache=true,cachedir=minted-cache]{minted}
\fi

\definecolor{GrayCodeBlock}{RGB}{241,241,241}
\definecolor{BlackText}{RGB}{110,107,94}
\definecolor{BlueTypename}{RGB}{17,86,182}
\definecolor{RedTypename}{RGB}{182,86,17}
\definecolor{GreenString}{RGB}{96,172,57}
\definecolor{PurpleKeyword}{RGB}{184,84,212}
\definecolor{GrayComment}{RGB}{170,170,170}
\definecolor{GrayNumber}{RGB}{200,200,200}
\definecolor{GoldDocumentation}{RGB}{180,165,45}
\ifarxiv
\else
\fi

\SetKwProg{CRDT}{crdt}{}{}
\SetKwInput{InitialState}{initialState}
\SetKwProg{Function}{function}{}{}
\SetKwProg{Merge}{merge}{}{}
\SetKwProg{Gossip}{gossip}{}{}
\SetKwProg{Operation}{operation}{}{}
\SetKwProg{Query}{query}{}{}
\SetKwBlock{Loop}{loop}{end}
\SetKw{Break}{break}
\SetKw{Send}{send}

\AtBeginDocument{%
  \providecommand\BibTeX{{%
    \normalfont B\kern-0.5em{\scshape i\kern-0.25em b}\kern-0.8em\TeX}}}

\newif\ifcomments
\commentstrue
\ifcomments
    \providecommand{\shadaj}[1]{{\protect\color{brown}{\bf [shadaj: #1]}}}
    \providecommand{\conor}[1]{{\protect\color{red}{\bf [conor: #1]}}}
    \providecommand{\alvin}[1]{{\protect\color{purple}{\bf [alvin: #1]}}}
    \providecommand{\mae}[1]{{\protect\color{blue}{\bf [mae: #1]}}}
    \providecommand{\joe}[1]{{\protect\color{teal}{\bf [joe: #1]}}}
    \providecommand{\jmh}[1]{{\protect\color{teal}{\bf [joe: #1]}}}
    \providecommand{\david}[1]{{\protect\color{green}{\bf [david: #1]}}}
    \providecommand{\chris}[1]{{\protect\color{violet}{\bf [chris: #1]}}}
    \providecommand{\davidmwei}[1]{{\protect\color{pink}{\bf [david wei: #1]}}}
    \providecommand{\kaushik}[1]{{\protect\color{orange}{\bf [kaushik: #1]}}}
    \providecommand{\justin}[1]{{\protect\color{green}{\bf [justin: #1]}}}
    \providecommand{\mingwei}[1]{{\protect\color{rhodamine}{\bf [mingwei: #1]}}}
    \providecommand{\rithvik}[1]{{\protect\color{red}{\bf [rithvik: #1]}}}
    \providecommand{\nc}[1]{{\protect\color{pink}{\bf [nc: #1]}}}
    \providecommand{\accheng}[1]{{\protect\color{olive}{\bf [accheng: #1]}}}
    \providecommand{\dan}[1]{{\protect\color{purple}{\bf [dan: #1]}}}
\else
    \providecommand{\shadaj}[1]{}
    \providecommand{\conor}[1]{}
    \providecommand{\alvin}[1]{}
    \providecommand{\mae}[1]{}
    \providecommand{\joe}[1]{}
    \providecommand{\jmh}[1]{}
    \providecommand{\david}[1]{}
    \providecommand{\chris}[1]{}
    \providecommand{\davidmwei}[1]{}
    \providecommand{\kaushik}[1]{}
    \providecommand{\justin}[1]{}
    \providecommand{\mingwei}[1]{}
    \providecommand{\rithvik}[1]{}
    \providecommand{\nc}[1]{}
    \providecommand{\accheng}[1]{}
    \providecommand{\dan}[1]{}
\fi

\begin{document}

\title{Suki: Choreographed Distributed Dataflow in Rust}


\author{Shadaj Laddad}
\affiliation{%
 \institution{UC Berkeley}\city{} \country{USA}}
 \email{shadaj@cs.berkeley.edu}

\author{Alvin Cheung}
\affiliation{%
 \institution{UC Berkeley}\city{} \country{USA}}
 \email{akcheung@cs.berkeley.edu}

\author{Joseph M. Hellerstein}
\affiliation{%
 \institution{UC Berkeley}\city{} \country{USA}}
 \email{hellerstein@cs.berkeley.edu}

\begin{abstract}
Programming models for distributed dataflow have long focused on analytical workloads that allow the runtime to dynamically place and schedule compute logic. Meanwhile, models that enable fine-grained control over placement, such as actors, make global optimization difficult. In this extended abstract, we present Suki, an embedded Rust DSL that lets developers implement streaming dataflow with explicit placement of computation. Key to this choreographic programming approach is our use of staged programming, which lets us expose a high-level Rust API while compiling local compute units into individual binaries with zero-overhead. We also explore how this approach, combined with Rust's trait system, enables a type-safe API for mapping dataflow programs to cloud computing resources.
\end{abstract}


\maketitle

\section{Introduction}
There is growing interest in better programming models for distributed systems --- ones that make it easier for developers to express their logic and reason about failures, that enable compilers to leverage advanced optimization techniques, and that better map to cloud computing architectures. Recent work, such as the Hydro Project~\cite{hydro}, has proposed \emph{streaming dataflow APIs} as an ideal interface for distributed programming. In this approach, distributed programs are implemented in terms of functional (map, filter, fold) and relational (join, cross product, set difference) operators. These operators have natural interpretations over streaming data, and are amenable to optimization~\cite{autocomp} since they reason about entire streams of data rather than elements.

A prominent example of this approach is Apache Flink~\cite{flink}, which offers a streaming dataflow API based on these principles. Flink comes with a powerful runtime, which enforces guarantees like exactly-once execution and dynamically balances computational load across a cluster of workers. This results in a relatively simple interface for developers, but comes at the cost of being \emph{too high level} for certain tasks. Because fault tolerance and compute placement are handled for the developer, there is no way to faithfully implement foundational protocols like Paxos or CRDT gossip.

On the flip side, languages like Bloom~\cite{bloom} focus on using dataflow APIs \emph{within} a single node. Programs written in this style look similar to actors, since they carry local state and explicitly interface with network inputs and outputs. But like actors, this approach syntactically scatters pieces of a distributed protocol according to \emph{where} they are run. This makes the programs harder to read and gets in the way of modularity.

In this extended abstract, we present our early work on combining the best of both worlds with Suki --- a choregraphic DSL for streaming dataflow. Key to our approach is the use of \emph{staged programming}~\cite{multi-stage-programming}, which separates the compilation into two phases: one that constructs a global dataflow graph and another that compiles the logic for each physical location into optimized Rust binaries. By leveraging staged programming, rather than relying on runtime scheduling, we are able to achieve \textbf{zero-overhead choreographic programming}, with individual binaries amenable to classic low-level optimizations such as autovectorization. Suki has been used to implement foundational protocols like Paxos with performance exceeding 50 kops/s out of the box.

Suki also explores a new frontier for choreographic programming: \textbf{mapping compute locations to concrete deployments on cloud infrastructure}. Our dataflow API allows developers to specify which cloud machines they would like to deploy their program to, with our system automatically handling provisioning and networking. We discuss how the Rust type system makes it possible to enforce type-safety at the deployment layer to avoid dangling machines and mismatched network connections.

\section{Staged, Choreographed Dataflow}
The core of Suki is a DSL for building streaming dataflow that lets developers explicitly place subgraphs on different machines. Suki dataflow graphs have two key concepts which give developers fine-grained control over placement:
\begin{itemize}
\item \textbf{Locations} capture \emph{where} (machine or process) a stream is being materialized
\item \textbf{Streams} capture an \emph{unbounded} sequence of elements \emph{at} a particular location
\end{itemize}

The entire API exposed by Suki is executed in a \emph{staged context}. That is, programs written with Suki first execute directly on the developer's laptop to generate the global dataflow graph. A second phase processes this graph to generate Rust sources for the subgraph projection at each location. Finally, these sources are compiled for the target deployment platform and can be launched to execute the computation.

\subsection{Locations and Specs}
The primary API that distinguishes Suki from existing streaming languages are \emph{locations}, which enable choreographic programming of dataflows that span across several nodes. To acquire a \texttt{Location} value in a Suki program, the developer must ask for a \emph{specification}, which captures the metadata necessary for deploying a concrete instance (Section~\ref{sec:deploy}). These specifications can be reused to create several locations with the same deployment logic.

The simplest type of location is a \textbf{process}, which represents \emph{exactly one} instance of the computation assigned to it. In Suki, we can construct a process from a \texttt{ProcessSpec}:

\begin{minted}{rust}
pub fn my_dataflow<'a, D: Deploy<'a>>(
    flow: &FlowBuilder<'a, D>,
    process_spec: &impl ProcessSpec<'a, D>
) {
    let my_process = flow.process(process_spec);
    ...
}
\end{minted}

Now, we can initialize a computation by placing it at our newly created location. For example, we can initialize a stream that generates elements from an iterable source. Here we see the first use of \emph{staged programming}: the expression for the iterable must be \textbf{quoted} using the \texttt{q!} macro so that it can be \textbf{spliced} into the final compiled program.

\begin{minted}{rust}
let numbers = flow.source_iter(&my_process, q!(0..5));
\end{minted}

Our approach for bringing staged programming to Rust also features some novel techniques, but is out of scope for this discussion. What is important here is that the contents of the \texttt{q!} macro are captured so they can be spliced into the generated code for the corresponding location, but are also typechecked in the staged program so that \texttt{numbers} has the type \texttt{Stream<u32>}.

\subsection{Dataflow Operators}
Now that we have initialized our stream at a particular location, we can perform computations with it! Suki provides all of the familiar functional and relational operators with a standard type-safe API. What make Suki's API somewhat unique is that all user-defined functions must be wrapped in \texttt{q!}, so that they can be spliced into generated code.

By default, all operators in Suki consume input streams that must all be at the same location, and return a new stream that is also at that location. For example, we can filter and transform the data from before:

\begin{minted}{rust}
let transformed_numbers = numbers.filter(q!(|v| v > 2)).map(q!(|v| v * 2));
\end{minted}

Under the hood, the implementations of these operators add to the global dataflow graph (stored in the \texttt{FlowBuilder}). During compilation, Suki slices the graph by locations and emits low-level Rust sources containing the runtime logic each subgraph. By leveraging staged programming, this implementation is relatively lightweight---we simply capture the tokens for each staged function and emit them (nearly) verbatim into the per-location sources.

\subsection{Networking}
So far, what we've seen looks like traditional dataflow programming, because none of our graphs have spanned multiple locations. To move streams between locations, Suki provides special \emph{networking} operators that correspond to using TCP (or similar) channels between physical machines.

To set the scene, let's create a second location, in this case reusing our existing specification:
\begin{minted}{rust}
let my_second_process = flow.process(process_spec);
\end{minted}

Now, we can choreograph our dataflow across these two locations by using the \texttt{send\_bincode} operator, which sends a stream to another location by using the \texttt{bincode} serialization library. This operator returns a new stream with the same element type as the original one, but with the location set to the target we passed in:

\begin{minted}{rust}
let on_second = transformed_numbers.send_bincode(&my_second_process);
\end{minted}

When generating per-location sources, these network operators are compiled sender and receiver logic on each end. We can apply operators to this stream which will be executed at the second location:

\begin{minted}{rust}
on_second.for_each(q!(|v| println!("{}", v)));
\end{minted}

With this simple abstraction for moving streams between locations, Suki can describe a wide range of distributed computing primitives, such as heartbeats. These building blocks can be represented in user code as separate Rust functions that take locations and specifications as parameters. This enables strong modularity along the lines of protocols, rather than splitting the code by location.

\subsection{Clusters and Partitioning}
So far, we have looked at static deployments where the exact number of compute units is known ahead of time. But in many distributed algorithms, such as MapReduce~\cite{mapreduce}, the dataflow graph contains pieces that can be dynamically scaled through partitioning and replication techniques. Suki supports such programs with a second type of location: \textbf{clusters}.

Clusters in Suki represent a SIMD-style programming model, where many instances of the same dataflow will be run on different pieces of data on different machines. This minimizes the API surface of Suki, since the exact same stream operators can be used even when the stream is assigned to a cluster location. What changes when interacting with clusters are the networking APIs. Without clusters, all communication was \emph{one-to-one}. But with clusters, we now have to consider the rest of the matrix: one-to-many, many-to-one, and many-to-many.

To let programs in Suki decide which members of a cluster a value should be sent to, and identify which cluster member was a sender, we introduce the concept of \textbf{cluster IDs}. These are unique identifiers for each member of a cluster that are \emph{only known at runtime}. Because Suki programs are staged, this means that user programs cannot interact with cluster IDs at the top-level---their usage is restricted to staged functions and operators.

When sending data to a cluster, the source stream must contain \texttt{(cluster ID, T)} data pairs. At runtime, each element is sent to the member with the matching ID. Similarly, when receiving data from a cluster, the recipient generates a stream of \texttt{(cluster ID, T)} pairs that mark which cluster member sent the value.

To access cluster membership information at runtime, Suki provides a special API (\texttt{ids()}) that returns a \emph{staged} set of cluster IDs. By loading this into a stream using \texttt{source\_iter}, patterns like broadcast become a simple cross-product between two streams:

\begin{minted}{rust}
pub fn my_broadcast<'a, D: Deploy<'a>>(
    flow: &FlowBuilder<'a, D>,
    process_spec: &impl ProcessSpec<'a, D>,
    cluster_spec: &impl ClusterSpec<'a, D>
) {
    let my_process = flow.process(process_spec);
    let my_cluster = flow.cluster(cluster_spec);

    let data_to_broadcast = flow.source_iter(&my_process, q!(0..5));

    let stream_of_cluster_ids = flow.source_iter(
      &my_process, // load the IDs at the process
      my_cluster.ids() // we are interested in membership of the cluster
    );
    
    stream_of_cluster_ids
      .cross_product(data_to_broadcast)
      .send_bincode(&my_cluster)
      .for_each(q!(|v| println!("{}", v)));
}
\end{minted}

Partitioning can be implemented in a similar manner, by mapping keys to cluster IDs with a hash function. Similarly, randomized CRDT gossip looks nearly identical to our broadcast program, but samples the cluster IDs rather than using the entire stream. The combination of processes and clusters yields significant expressive power. In our experiments, we have implemented several complex distributed protocols, including Paxos, using just the cluster interface.

\section{Type-Safe Deployments}
\label{sec:deploy}

Implementing a distributed program is just one half of the story; deploying it is equally complex. Especially in the world of cloud computing providers and multi-cloud architectures, it is easy to make mistakes such as misconfiguring network ports or launching unnecessary virtual machines. We believe that deployments are \emph{just as important} to the success of choreographic programming as the computational layer, since developers need to confidently update subcomponents without worrying about their deployment logic going out-of-sync.

Suki aims to solve this by providing an integrated layer for mapping the \emph{virtual} locations in a dataflow program to \emph{physical} deployment plans in the form of Terraform specifications. By generating the deployment logic \emph{from} the dataflow graph, developers can rest assured that their program will execute exactly as they intended, while preserving the ability to make decisions about machine types and datacenter selection.

Let us return to the broadcast example from before. When deploying this program, we need to specify which cloud machines the process and cluster should run on. Suki enforces this through the type-system, by requiring us to provide cloud machine specifications as arguments when invoking the \texttt{my\_broadcast} function. In this case, we map each process to a single Google Cloud instance, and each cluster to a pair of instances:

\begin{minted}{rust}
let flow = FlowBuilder::new();
my_broadcast(
    &flow,
    &DeployProcessSpec::new(|| {
        GCPComputeEngineHost::new(
            "e2-micro", // machine type
            "debian-cloud/debian-11", // image
            "us-west1-a" // region
        )
    }),
    &DeployClusterSpec::new(|| {
        (0..2).map(|_| {
            GCPComputeEngineHost::new(
                "e2-micro",
                "debian-cloud/debian-11",
                "us-west1-a"
            )
        }).collect()
    })
);
\end{minted}

Because we have separate types that implement the \texttt{ProcessSpec} and \texttt{ClusterSpec} traits, we can enforce type-safety by ensuring each location has a corresponding machine in the deployment. Every time a specification is instantiated into a process or cluster, Suki automatically adds the GCP parameters to a Terraform configuration for the global deployment. Suki also provides runtime APIs that evaluate the Terraform config, deploy the Rust binaries compiled for each location, and perform service discovery to connect the binaries according to the choreographed network topology.

By leveraging multiple location specifications, developers can build dataflow programs that respect modern constraints such as data sovereignty laws. For example, a Suki program can take separate cluster specs for locations in the US and Europe, so that the choreographed program respects data locality regulations. These specs can then be fulfilled by cloud machines in the respective regions, ensuring that the placement of data in the Suki program matches execution.

This integration also enables the ability to \emph{optimize} deployments. For example, we can automatically choose which region each cloud machine is placed in based on the network connections between the corresponding locations. Similarly, members of a cluster can be automatically distributed across several cloud providers and regions to satisfy fault tolerance constraints. We believe that such techniques, combined with optimizations over the dataflow graph, point towards co-design of dataflow and deployment.

\section{Future Work}
\subsection{Optimizers over the Suki IR}
One of the key goals of Suki is to support optimizers that reason about the global dataflow graph, rather than per-node logic. When using Suki, the \texttt{FlowBuilder} stores the dataflow graph in a global IR that captures both the core operators executed at each location as well as networking operators between them. Rewrites over this IR can be very powerful, as they can move computations \emph{between} locations or fuse logic that is initially placed on separate machines.

A particularly intriguing set of optimizations over the Suki IR are those that introduce \emph{new} locations to increase the performance of the dataflow. Recent work has shown that pipelining and partitioning can be used to automatically increase the throughput of classic coordination protocols~\cite{autocomp}, and we believe that the Suki IR can be a great fit for implementing such optimizers. Furthermore, since Suki integrates deeply with deployment mechanisms, it may be possible to perform optimization over live applications by dynamically replacing the subgraphs for different locations. We are actively implementing these optimizations using techniques like e-graphs~\cite{egg} to explore the large space of program rewrites.

\subsection{Interpreters over an Untyped Interface}
Currently, Suki only exposes a strongly-typed interface for streams. This requires the type of each stream's elements to be known \emph{ahead of time}. This is great for application developers, as it ensures that the code generated for each location will compile successfully. But it can get in the way of programs that want to dynamically generate streams of different types.

An interesting example of this limitation is implementing compilers for \emph{other languages} by leveraging Suki. For example, a Datalog engine can be implemented in terms of Suki's built-in operators. But if the input is a string of Datalog code, there is no way to know what the types of the underlying streams will be in the staged dataflow logic.

While strongly typed streams are useful, they are in no way \emph{necessary} for Suki's choreographic programming interface. Indeed, we can modify the operators to capture staged snippets of code that may involve dynamically selected types by deferring typechecking to the generated low-level Rust sources. We are excited for this approach to make it easier to construct compilers that target distributed systems by "interpreting" the source programs with a Suki implementation.

\section{Conclusion}
Many modern applications involve distributed streaming logic that reacts to live data with low latency, but existing programming models for such systems force a tradeoff between modularity and fine-grained placement. With choreographic programming, we have the opportunity to fill the gap with languages that allow programs spanning machines to be written in one place. The Suki DSL offers a step in this direction by using staged programming to provide choreographic streaming APIs with zero runtime cost. Furthermore, Suki expands the scope of choreographic programming to include interactions with cloud providers, which enables safe and efficient deployments.

\begin{acks}
We thank our anonymous reviewers for their insightful feedback on this paper. This work is supported in part by National Science Foundation CISE Expeditions Award CCF-1730628, IIS-1955488, IIS-2027575, GRFP Award DGE-2146752, DOE award DE-SC0016260, ARO award W911NF2110339, and ONR award N00014-21-1-2724, and by gifts from Astronomer, Google, IBM, Intel, Lacework, Microsoft, Mohamed Bin Zayed University of Artificial Intelligence, Nexla, Samsung SDS, Uber, and VMware.
\end{acks}

\bibliographystyle{ACM-Reference-Format}
\bibliography{bibliography}


\begin{thebibliography}{7}


\ifx \showCODEN    \undefined \def \showCODEN     #1{\unskip}     \fi
\ifx \showDOI      \undefined \def \showDOI       #1{#1}\fi
\ifx \showISBNx    \undefined \def \showISBNx     #1{\unskip}     \fi
\ifx \showISBNxiii \undefined \def \showISBNxiii  #1{\unskip}     \fi
\ifx \showISSN     \undefined \def \showISSN      #1{\unskip}     \fi
\ifx \showLCCN     \undefined \def \showLCCN      #1{\unskip}     \fi
\ifx \shownote     \undefined \def \shownote      #1{#1}          \fi
\ifx \showarticletitle \undefined \def \showarticletitle #1{#1}   \fi
\ifx \showURL      \undefined \def \showURL       {\relax}        \fi
\providecommand\bibfield[2]{#2}
\providecommand\bibinfo[2]{#2}
\providecommand\natexlab[1]{#1}
\providecommand\showeprint[2][]{arXiv:#2}

\bibitem[Alvaro et~al\mbox{.}(2011)]%
        {bloom}
\bibfield{author}{\bibinfo{person}{Peter Alvaro}, \bibinfo{person}{Neil
  Conway}, \bibinfo{person}{Joseph~M. Hellerstein}, {and}
  \bibinfo{person}{William~R. Marczak}.} \bibinfo{year}{2011}\natexlab{}.
\newblock \showarticletitle{Consistency Analysis in Bloom: a {CALM} and
  Collected Approach}. In \bibinfo{booktitle}{\emph{Fifth Biennial Conference
  on Innovative Data Systems Research, {CIDR} 2011, Asilomar, CA, USA, January
  9-12, 2011, Online Proceedings}}. \bibinfo{pages}{249--260}.
\newblock
\urldef\tempurl%
\url{http://cidrdb.org/cidr2011/Papers/CIDR11\_Paper35.pdf}
\showURL{%
\tempurl}


\bibitem[Carbone et~al\mbox{.}(2015)]%
        {flink}
\bibfield{author}{\bibinfo{person}{Paris Carbone}, \bibinfo{person}{Asterios
  Katsifodimos}, \bibinfo{person}{Stephan Ewen}, \bibinfo{person}{Volker
  Markl}, \bibinfo{person}{Seif Haridi}, {and} \bibinfo{person}{Kostas
  Tzoumas}.} \bibinfo{year}{2015}\natexlab{}.
\newblock \showarticletitle{Apache flink: Stream and batch processing in a
  single engine}.
\newblock \bibinfo{journal}{\emph{The Bulletin of the Technical Committee on
  Data Engineering}} \bibinfo{volume}{38}, \bibinfo{number}{4}
  (\bibinfo{year}{2015}).
\newblock


\bibitem[Cheung et~al\mbox{.}(2021)]%
        {hydro}
\bibfield{author}{\bibinfo{person}{Alvin Cheung}, \bibinfo{person}{Natacha
  Crooks}, \bibinfo{person}{Joseph~M Hellerstein}, {and} \bibinfo{person}{Mae
  Milano}.} \bibinfo{year}{2021}\natexlab{}.
\newblock \showarticletitle{New directions in cloud programming}.
\newblock \bibinfo{journal}{\emph{The Conference on Innovative Data Systems
  Research (CIDR)}}, \bibinfo{numpages}{14}~pages.
\newblock


\bibitem[Chu et~al\mbox{.}(2024)]%
        {autocomp}
\bibfield{author}{\bibinfo{person}{David C.~Y. Chu}, \bibinfo{person}{Rithvik
  Panchapakesan}, \bibinfo{person}{Shadaj Laddad}, \bibinfo{person}{Lucky~E.
  Katahanas}, \bibinfo{person}{Chris Liu}, \bibinfo{person}{Kaushik
  Shivakumar}, \bibinfo{person}{Natacha Crooks}, \bibinfo{person}{Joseph~M.
  Hellerstein}, {and} \bibinfo{person}{Heidi Howard}.}
  \bibinfo{year}{2024}\natexlab{}.
\newblock \showarticletitle{Optimizing Distributed Protocols with Query
  Rewrites}.
\newblock \bibinfo{journal}{\emph{Proc. ACM Manag. Data 2, N1 (SIGMOD)}},
  Article \bibinfo{articleno}{2} (\bibinfo{date}{Feb.} \bibinfo{year}{2024}),
  \bibinfo{numpages}{25}~pages.
\newblock
\urldef\tempurl%
\url{https://doi.org/10.1145/3639257}
\showDOI{\tempurl}


\bibitem[Dean and Ghemawat(2008)]%
        {mapreduce}
\bibfield{author}{\bibinfo{person}{Jeffrey Dean} {and} \bibinfo{person}{Sanjay
  Ghemawat}.} \bibinfo{year}{2008}\natexlab{}.
\newblock \showarticletitle{MapReduce: simplified data processing on large
  clusters}.
\newblock \bibinfo{journal}{\emph{Commun. ACM}} \bibinfo{volume}{51},
  \bibinfo{number}{1} (\bibinfo{date}{jan} \bibinfo{year}{2008}),
  \bibinfo{pages}{107–113}.
\newblock
\showISSN{0001-0782}
\urldef\tempurl%
\url{https://doi.org/10.1145/1327452.1327492}
\showDOI{\tempurl}


\bibitem[Taha and Sheard(1997)]%
        {multi-stage-programming}
\bibfield{author}{\bibinfo{person}{Walid Taha} {and} \bibinfo{person}{Tim
  Sheard}.} \bibinfo{year}{1997}\natexlab{}.
\newblock \showarticletitle{Multi-stage programming with explicit annotations}.
\newblock \bibinfo{journal}{\emph{SIGPLAN Not.}} \bibinfo{volume}{32},
  \bibinfo{number}{12} (\bibinfo{date}{dec} \bibinfo{year}{1997}),
  \bibinfo{pages}{203–217}.
\newblock
\showISSN{0362-1340}
\urldef\tempurl%
\url{https://doi.org/10.1145/258994.259019}
\showDOI{\tempurl}


\bibitem[Willsey et~al\mbox{.}(2021)]%
        {egg}
\bibfield{author}{\bibinfo{person}{Max Willsey}, \bibinfo{person}{Chandrakana
  Nandi}, \bibinfo{person}{Yisu~Remy Wang}, \bibinfo{person}{Oliver Flatt},
  \bibinfo{person}{Zachary Tatlock}, {and} \bibinfo{person}{Pavel Panchekha}.}
  \bibinfo{year}{2021}\natexlab{}.
\newblock \showarticletitle{Egg: Fast and Extensible Equality Saturation}.
\newblock \bibinfo{journal}{\emph{Proc. ACM Program. Lang.}}
  \bibinfo{volume}{5}, \bibinfo{number}{POPL}, Article \bibinfo{articleno}{23}
  (\bibinfo{date}{jan} \bibinfo{year}{2021}), \bibinfo{numpages}{29}~pages.
\newblock
\urldef\tempurl%
\url{https://doi.org/10.1145/3434304}
\showDOI{\tempurl}


\end{thebibliography}

\end{document}